\documentclass[12pt]{iopart}

\usepackage{iopams}  
\usepackage{graphicx}

\usepackage{dcolumn}
\usepackage{bm}
\usepackage{cite}

\usepackage{rotating}
\usepackage{array}
\usepackage{amssymb}

\def\v#1{{\bf#1}}
\def\be{\begin{equation}}
\def\ee{\end{equation}}
\def\bea{\begin{eqnarray}}
\def\eea{\end{eqnarray}}

\newtheorem{theorem}{Proposition}

\def\scal{\mbox{$\cal S\,$}}
\def\ucal{\mbox{$\cal U\,$}}
\def\<{\langle}
\def\>{\rangle}

\begin{document}
\nocite{*}

\title{Inverse lattice design and its application to bent waveguides}

\author{E. Rivera-Moci\~nos and E. Sadurn\'i}
\address{Instituto de F\'isica, Benem\'erita Universidad Aut\'onoma de Puebla,
Apartado Postal J-48, 72570 Puebla, M\'exico}
\ead{erivera@ifuap.buap.mx, sadurni@ifuap.buap.mx}

\begin{abstract}
This paper is divided in two parts. In the first part, the inverse spectral problem for tight-binding hamiltonians is studied. This problem is shown to have an infinite number of solutions for properly chosen energies. The space of such solutions is characterized by a hypersurface in the space of hopping amplitudes (i.e. couplings), whose dimension is half the number of sites in the array. Low dimensional examples for short chains are carefully studied and a table of exactly solvable inverse problems is provided in terms of Lie algebraic structures. With the aim of providing a method to generate lattice configurations, a set of equations for coupling constants in terms of energies is obtained; this is done by means of a new formula for the calculation of characteristic polynomials. Two examples with randomly generated spectra are studied numerically, leading to peaked distributions of couplings. In the second part of the paper, our results are applied to the design of bent waveguides, reproducing specific spectra below propagation threshold. As a demonstration, the Dirac and the finite oscillator are realized in this way. A few partially isospectral configurations are also presented.
\end{abstract}



\section{Introduction}

The continuous interest in inverse spectral theory \cite{band2009, kac1966} has reached many important results connected with cavities, billiards and their inherent quantum chaology \cite{berry1989, stoeckmann1999}. Apart from the purest interest of Kac's question and the shape of drums, fresh applications of quantum mechanics seem to draw the attention to similar problems in different areas such as propagation in crystals. Recent examples in experiments connected with the quantum behaviour of matter \cite{bloch2005, oberthaler1996} illustrate the important role of tight-binding lattices and their shapes. This is also true for other emulations of solids in mesoscopic physics, which sacrifice quantumness in favour of geometrical flexibility, with the aim of exploring more and more effects inexpensively \cite{laurent2007, sadurni2010-1, sadurni2013, franco2013}. Along with these examples comes the necessity of dealing with inverse spectral problems defined on lattices of any dimensionality, in particular tight-binding chains. Thus, inverse lattice design can be conceived as an answer to the problem of a discrete shape producing a specific spectral feature. 

\begin{figure}[h!]
\begin{center}  \includegraphics[width=12cm]{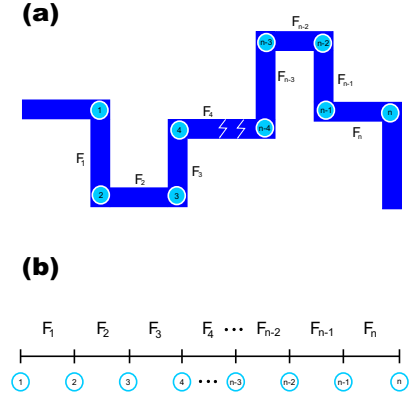} \end{center}
\caption{\label{fig:I.0} Systems described in this paper: bent waveguides with multiple corners and tight-binding arrays of resonators.}
\end{figure}

The incarnations of tight-binding models are of a great variety, but the focus of this paper shall be centered in waveguides, since they can be regarded as elongated variants of the more traditional drum. See figure \ref{fig:I.0}. Waveguides can also be open at their ends, which makes them accessible through scattering, with slight perturbative effects in their resonating frequencies when compared with their closed counterparts. The question is, can we hear the shape of a pipe? 

In this paper we provide a general answer to the following two practical questions: Is it possible to produce an arbitrary finite spectrum by the mere use of a one-dimensional chain {\it without\ }local potentials? Are these results applicable to the transport properties of a scattering device? The answers shall be delivered by characterizing the space of all tight-binding hamiltonians that produce a suitably defined spectrum, together with their plausible realization in bent waveguides operating with a well-established binding mechanism at corners \cite{exner1989, schult1989, sadurni2010-2, bittner2013}.   

Similar problems have been proposed in previous works; notably, the system of interest has a close resemblance with quantum graphs, for which there are extensive studies \cite{kuchment2004, kuchment2005, kurasov2005}. However, there are also sharp differences between our waveguides and one-dimensional wires conforming a graph. For instance, evanescent transport is possible only in the presence of a non-vanishing thickness, and it is through this mechanism that we shall establish finite difference equations. In this respect, it is important to mention that significant results in the field of graphs can be found in \cite{pankrashkin2006}, relating the spectrum of a continuous one-dimensional Laplacian with that of a finite difference operator defined on graph vertices. Nevertheless, the design of a specific graph producing an arbitrary spectrum is rarely given in the literature.

We present our discussion as follows. In section \ref{sec:2} we characterize the spaces of hamiltonians for a given spectrum and we generate a set of algebraic equations that relate hopping amplitudes (couplings) in terms of energies as parameters. This part is of a mathematical nature and it is carefully presented. 
As an application, in section \ref{sec:3} we study bent waveguide realizations of tight-binding arrays. 
We demonstrate that exactly solvable systems of the inverse problem can be emulated by explicit design of the lattice: the Dirac oscillator, the finite oscillator and partially isospectral chains are presented. We conclude in section \ref{sec:4}.

\section{The problem of inverse design in nearest neighbour models \label{sec:2}}

We shall work with nearest neighbour tight-binding systems of identical sites, which belong to the wider class of Jacobi matrices. We must mention that previous work such as \cite{gesztezy1997} and references therein, deals with the inverse spectral theory of such operators. Although the theory has more general hypotheses related to all posible tridiagonal matrices, the claimed solution in such references rests on the assumption of two known spectra for two Hamiltonians belonging to the same sequence -- see Corollary 2.5 in \cite{gesztezy1997}. The intertwining of levels and the specific form of the spectral function allow to determine univocally the elements of one of the Hamiltonians (either the first or second in the sequence). However, our inverse problem is restricted to the knowledge of only one spectrum. We shall see that this explains the existence of infinite solutions. The determination of such a space of solutions can be seen as a complementary result to the known literature. 

Let us establish our mathematical notation: We choose units such that $\hbar=1$, we denote by $N+1$ the number of sites in the array and we assume that any state $|\phi\>$ can be written as a superposition of a complete set of localized (atomic) states $|n\>$:

\bea
|\phi \> = \sum_{n=1}^{N+1} \phi_n |n\>.
\label{1.1}
\eea 
We consider the wavefunction $\< x|n \>$ to be concentrated around the site located at $x_n$ (e.g. a Wannier function \cite{wannier1937, marzari2012}, but without band index for simplicity). Since we deal with arrays of identical sites, we must have identical localized functions related by translations $\<x | n-1\> = \<x-x_n+x_{n-1} | n\>$. The atomic states themselves $|n\>$ can be translated by means of a unitary operator $T$

\bea
T |n\> = |n-1\>, \qquad T^{\dagger} |n\> = |n+1\>.
\label{1.2}
\eea
The matrix elements of $T$ are clearly given by $\<n|T|m\> =\delta_{n,m-1}$. We can define an observable $\hat N$ associated to the site label $n$ with the property

\bea
\hat N |n\>=n |n\>, \qquad \<n|\hat N|m \>=n \delta_{n,m}.
\label{1.3}
\eea
For any function $F$ of the operator $\hat N$ alone, we must have

\bea
F(\hat N) |n\>=F(n) |n\> \equiv F_n |n\>.
\label{1.4}
\eea
The stationary Schr\"odinger equation corresponding to nearest neighbour chains with variable couplings but without local potentials is thus

\bea
H |\phi_k \> = E_k |\phi_k \> = \left[ F(\hat N) T + T^{\dagger} F(\hat N) \right] |\phi_k \>, \quad 1 \leq k \leq N+1,
\label{1.5}
\eea
where now $F(\hat N)$ encodes the information on the variable couplings, according to their location in the chain. In fact, (\ref{1.5}) can be written as a recurrence  if we set

\bea
|\phi_k\> = \sum_{n=1}^{N+1} \phi_n^{k} |n\>,
\label{1.6}
\eea 
from which the following relation holds:

\bea
E_k \phi_n^k = F_n \phi_{n+1}^k + F_{n-1}^* \phi_{n-1}^k, \quad 1 \leq k \leq N+1, \quad 1 \leq n \leq N+1.
\label{1.7}
\eea
Since the system is finite, the previous relation is supplemented with fictitious boundary conditions $\phi_0=\phi_{N+2}=0$. Similarly, the couplings to fictitious sites must vanish $F_0 = F_{N+1}=0$. 

With these conditions, the physical block of the system is determined by the non-vanishing elements of the operator $F(\hat N)$. Since we are interested in completely connected chains, we impose $F_n \neq 0$ for $1 \leq n \leq N$, but it is important to stress that $N$ can be arbitrarily large.

A few results can be established already: The spectrum $\sigma(H)$ -- with $H$ as in (\ref{1.5})-- is not quite arbitrary.

\begin{theorem} If $E_k \in \sigma(H)$, then $-E_k \in \sigma(H)$, i.e. $\sigma(H)$ is symmetrical around the origin. \end{theorem} To prove this assertion, we multiply by $(-1)^{n+1}$ both sides of (\ref{1.7}) and define $\psi_{n}^k =(-1)^n \phi_n^k$, leading to

\bea
 - E_k \psi_n^k = F_n \psi_{n+1}^k + F_{n-1}^* \psi_{n-1}^k,
\label{1.8}
\eea
which implies that $|\psi_k\> = \sum_n \psi_n^k|n\>$ is an eigenstate of $H$ with eigenvalue $- E_k$ $\square$. 

A simple corollary of this result is that for $N$ even, $0 \in \sigma(H)$, i.e. at least one $E_k$ should vanish.

Another result of interest here is the nature of wavefunctions in connection with couplings. \begin{theorem} Let $\tilde H = |F(\hat N)| T + T^{\dagger} |F(\hat N)|$ with $|F(\hat N)| |n\> \equiv |F_n| |n\>$. Then $\sigma(H) = \sigma(\tilde H)$. \end{theorem} The proof is given by a gauge transformation when $F_n$ is allowed to be complex. If we set $F_n = e^{i \delta_n}|F_n|$, $\Delta_n \equiv \sum_{j=1}^{n-1} \delta_j$ and multiply both sides of (\ref{1.7}) by $e^{i \Delta_n}$, we find

\bea
E_k ( e^{i \Delta_n}\phi_n^k) = F_n (e^{i \Delta_{n+1}} \phi_{n+1}^k) + F_{n-1}^* (e^{i \Delta_{n-1}}\phi_{n-1}^k).
\label{1.9}
\eea
From this relation it follows that the new state 

\bea
|\psi_k\> \equiv \sum_{n=1}^{N+1} e^{i \Delta_{n}} \phi_{n}^k | n \>
\label{1.10}
\eea
has energy $E_k$ with hamiltonian $\tilde H$ $\square$. 

Therefore, without loss of generality, we can work with real and positive couplings $F_n$ in one-dimensional systems without loops. Although we are not concerned now with a non-trivial topology of the chain, it is worth mentioning that the influence of magnetic fields would destroy (\ref{1.9}).

In passing, we note that in certain applications related to transmission of information along chains (e.g. bosonic or fermionic matter in optical lattices \cite{bloch2005, oberthaler1996}) one needs a second-quantization scheme. The formalism previously developed can be generalized easily to such situations: The local-field replacements $\phi_n \mapsto \hat a_n$, $\phi_n^* \mapsto \hat a^{\dagger}_n$, with canonical relations $\left[ \hat a_n , \hat a^{\dagger}_m \right]=\delta_{n,m}$  or $\left\{ \hat a_n , \hat a^{\dagger}_m \right\}=\delta_{n,m}$ lead to spectral problems similar to (\ref{1.7}).

Returning to our problem as stated in (\ref{1.5}), we recognize that $H$ can be identified with the position operator $X$ of an $f$- or $q$- deformed oscillator (e.g. deformations of Heisenberg algebras {\it \`a la\ }Man'ko \cite{biedenharn1989, macfarlane1989, manko1993}). This type of quantum optical application requires an operator $\hat N -1 $ representing the occupation number of photons at a given frequency, i.e. a normal mode in a cavity. What we note from this connection is that indeed, position operators have symmetric spectra about the origin and that eigenvalues can be distributed either continuously, discretely, in bounded regions, in closed intervals, or in the whole real line.

Finally, the inverse spectral problems that we tackle in this paper must be fed \footnote{An exception may arise when one of the eigenfunctions of $H$ is known from the outset. We touch upon this case in other sections.}with the following information:
\begin{itemize}
\item[ i)] A symmetric, but otherwise arbitrary set $\sigma(H)$, \item[ii)] A set of real coupling constants $F_n$, \item[iii)] An arbitrary but fixed size $N+1$.
\end{itemize}
With this in mind, we characterize the spaces of solutions in what follows.

\subsection{The many solutions of the inverse spectral problem for variable couplings \label{sec:2.1}}

\begin{figure}

\begin{center}

%
%
%
%
%
%
%
%
%
%
%
%
%
%
%
%

\includegraphics[width=15cm]{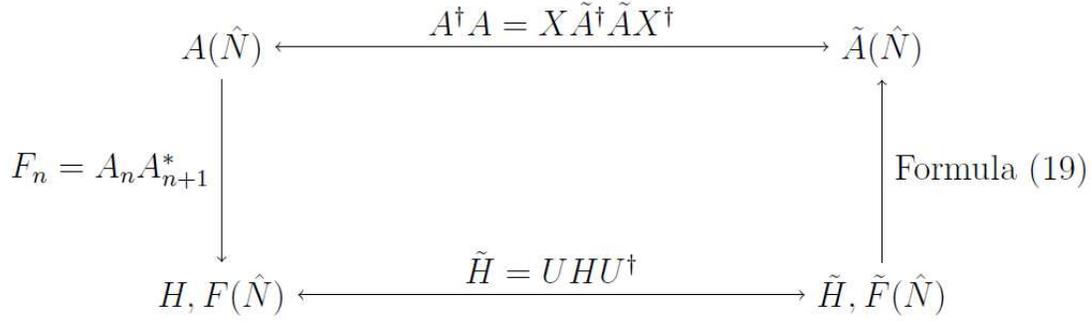}

\end{center}
\caption{Diagram showing the maps between isospectral tridiagonal hamiltonians (lower row) and their characterization in terms of $X$ (upper row).}
\label{fig:1}
\end{figure}

In this section we determine the space $\scal$ of all possible nearest-neighbour hamiltonians that lead to the same spectrum - complying, of course, with the requirements of the previous section. We shall see that such spaces can be characterized by orbits generated by linear operators, which however do not correspond to the whole space of $(N+1) \times (N+1)$ unitary matrices due to the nearest-neighbour restriction. We shall see as well that the gauge transformations defined by (\ref{1.10}) are but a trivial part of $\scal$. For a positive integer $n$, let us denote by $\ucal(n)$ the group of unitary $n\times n$ matrices and let us define $U_H(n)=\left\{ UHU^{\dagger} | U \in \ucal(n)\right\}$. Then, for a fixed $\sigma(H)$ we have an equivalence class $\scal_H$ containing $H$ such that

\bea
\left[ U_H(1) \right]^{N+1} \subset \scal_H \subset U_H (N+1)
\label{1.11}
\eea
where the contentions are proper except for $N=0$ or vanishing couplings. The process by which we shall work consists in the introduction of two auxiliary operators $A$ and $X$; $A$ encodes the information of the couplings and $X$ connects all the hamiltonians isospectral to the homogeneous chain. This is shown diagramatically in figure \ref{fig:1} and we proceed to explain it.

First, we determine the conditions for a set of couplings that lead to a certain $\sigma(H)$. Suppose that $\tilde H$ satisfies $\sigma(H)=\sigma(\tilde H)$; then there must be a unitary matrix $U$ such that:

\bea
U \left\{ F(\hat N) T + T^{\dagger} F^{\dagger}(\hat N) \right\} U^{\dagger} =  \tilde F(\hat N) T + T^{\dagger} \tilde F^{\dagger}(\hat N).
\label{1.12}
\eea
Now, we can prove that each hamiltonian is susceptible of a factorization

\bea
H= A(\hat N) H_0 A^{\dagger}(\hat N), \qquad  \tilde H= \tilde A(\hat N) H_0 \tilde A^{\dagger}(\hat N), \qquad H_0= T+T^{\dagger},
\label{1.13}
\eea
reflecting the fact that couplings can be obtained by lateral transformations acting on the homogenous chain hamiltonian $H_0$ (if the chain is finite, the system is sometimes referred to as a locally periodic structure \cite{griffiths2000}). The operator $A(\hat N)$ is constructed by observing that

\bea
A(\hat N) H_0 A^{\dagger}(\hat N) = A(\hat N) A^{\dagger}(\hat N + 1) T + T^{\dagger} A(\hat N + 1) A^{\dagger}(\hat N),
\label{1.14}
\eea
leading to $F(\hat N) = A(\hat N) A^{\dagger}(\hat N +1)$, and from here we can find the elements of $A$ using the recurrence

\bea
A_n A^*_{n+1} = F_n
\label{1.15}
\eea
or in polar form (i.e. $A_n \equiv |A_n| e^{i a_n}, F_n \equiv |F_n| e^{i f_n}$)
\bea
|F_n|= |A_n| |A_{n+1}|, \qquad f_n = a_n - a_{n+1}. 
\label{1.16}
\eea
The solutions are

\bea
a_n = \sum_{j=0}^{n-1} f_j, \qquad f_0 \in \Re
\label{1.17}
\eea
and
\bea
|A_{n+1}| = \frac{\prod_{k=0}^{\left[ | n/2 |\right]} |F_{n-2k}|  }{ \prod_{k=0}^{\left[ | n/2 |\right]} |F_{n-2k-1}| } \times A_1^{(-)^n}, \qquad 1 \leq n \leq N, \quad A_1 \in \Re_+ 
\label{1.18}
\eea
Furthermore, since $F_n$ can be chosen real and positive, we may work with 
\bea
A_{n+1} = \frac{\prod_{k=0}^{\left[ | n/2 |\right]} F_{n-2k}  }{ \prod_{k=0}^{\left[ | n/2 |\right]} F_{n-2k-1} } \times A_1^{(-)^n}, \qquad 1 \leq n \leq N, \quad A_1 \in \Re \backslash \{ 0 \}
\label{1.19}
\eea
and $a_n \equiv 0$. Now that we have constructed $A$ from any given $F$, we can also define its {\it inverse\ }restricting ourselves to the physical block, i.e. $A^{-1}A = A A^{-1} = \v 1_{N+1}$, with $\v 1_{N+1}$ the $ (N+1)\times (N+1) $ identity matrix. With this, we can transform (\ref{1.12}) into

\bea
H_0 = X H_0 X^{\dagger}, \qquad X \equiv \tilde A^{-1}(\hat N) U A(\hat N).
\label{1.20}
\eea
Our task is now simplified to the determination of all matrices $X$ subject to (\ref{1.20}). In other words, it is enough to analyze the locally periodic chain.

\subsubsection{The matrix $X$} The solutions of (\ref{1.20}) are multifold. We must note that $X$ could be non-unitary, which renders a solution space beyond the set of all symmetries of the periodic chain. We also note that the validity of (\ref{1.20}) is restricted to the physical $N+1$ dimensional block, which is more conveniently described by $N+1$ dimensional projectors $P$ as follows:

\bea
X P (T+T^{\dagger}) P^{\dagger} X^{\dagger} = P (T+T^{\dagger}) P^{\dagger}.
\label{1.21}
\eea
In order to find $X$, we compute the matrix elements of our operators in the eigenbasis of $P (T+T^{\dagger}) P^{\dagger}$. For infinite systems, the states in question are Bloch waves. For finite systems, both wavefunctions $ \< m | k \>$ and eigenvalues $E_k$ of $H_0$ are known (in fact, they are related to the Dirichlet kernel). We have

\bea
\sum_{n,m} \< k | X | m\> \< m | P (T+T^{\dagger}) P^{\dagger}    | n \> \< n | X^{\dagger} | k' \> = E_k \delta_{k,k'}
\label{1.22}
\eea
which means that the matrix elements of $X$ in mixed bases $X_{k; m} = \< k | X |m \>$ constitute an operator that diagonalizes $H_0$. Therefore, we must have

\bea
X_{k;m} = e^{-i \alpha(k)} \< k | m \>, \qquad X^*_{m; k} = e^{i \alpha(k)} \< m|k \>.
\label{1.23}
\eea
Here we have a freedom of choice, since the phase encoded by $\alpha(k)$ is so far arbitrary \footnote{In the limit $N \rightarrow \infty$, these phases are linked to the problem of the most localized Wannier functions. In linear arrays, this problem has been solved \cite{marzari2012}.}. This possibility renders $X$ unitary, but more general choices are attainable such as $\left[X, P(T+T^{\dagger})P^{\dagger} \right]=0$ and $(XX^{\dagger}-1)P(T+T^{\dagger})P^{\dagger}=0$, leading to

\bea
XX^{\dagger} = 1 + P_0,  \qquad P_0 (T+T^{\dagger}) = (T+T^{\dagger})P_0 = 0,
\label{1.24}
\eea
which is indeed more general than a unitarity condition. In fact, the projector $P_0$ onto the null space of $H_0$ exists whenever $N$ is even, making $N+1$ odd and at least one eigenvalue would vanish. For our purposes (general $N$), the choice (\ref{1.23}) is a reasonable one, as it contains a sufficient number of real parameters $\alpha(1), \alpha(2), ..., \alpha(N+1)$.

In addition to (\ref{1.21}), we have the condition $U=\tilde A X A^{-1}$ with $U$ unitary. This implies

\bea
(A^{\dagger})^{-1} X \tilde A^{\dagger} \tilde A X^{\dagger} A^{-1} = \v 1_{N+1}
\label{1.25}
\eea
or
\bea
  X \tilde A^{\dagger} \tilde A X^{\dagger} = A^{\dagger} A.
\label{1.26}
\eea
From here we see that given a matrix $\tilde A^{\dagger} \tilde A$, we can build all possible $A^{\dagger} A$ by choosing a specific $X$, which in turn amounts to a choice of the phases $\alpha(k)$. This expression can be given also in terms of matrix elements:

\bea
 |A_n|^2 \delta_{n,n'} &=& \sum_{k,k',m,m'} \<n|X|k\> \< k | m \> \< m |  \tilde A^{\dagger} \tilde A  | m'\> \< m' | k' \> \<k' | X^{\dagger} | n' \>  \nonumber \\
&=& \sum_{k,k',m} e^{i(\alpha(k')-\alpha(k))} \< n | k \> \< k | m \> \< m | k' \> \< k' | n' \> |\tilde A_m|^2 .
\label{1.27}
\eea
The diagonal part of (\ref{1.27}) represents indeed the orbits of a vector with components $|\tilde A_m|^2, m=1,...,N+1$ under the action of a linear operator that depends on the parameters $\alpha(k)$. On the other hand, the off-diagonal part restricts the possible $\alpha$'s as functions of $\tilde A$. This establishes that only some pairs $A, \tilde A$ can be related according to restrictions such as 

\bea
\sum_{k,k',m} \sin \left(\alpha(k')-\alpha(k) \right) \< n | k \> \< k | m \> \< m | k' \> \< k' | n' \> |\tilde A_m|^2 = 0,
\label{1.28}
\eea
\bea
|\sum_n |A_n|^2 \< k| n \> \< n| k' \>|^2 = | \sum_m |\tilde A_m|^2 \< k| m \> \< m| k' \>|^2,
\label{1.30}
\eea
\bea
\prod_{n=1}^{N+1} |\tilde A_n|^2 = \prod_{n=1}^{N+1} | A_n|^2,
\label{1.31}
\eea
which can be easily deduced. 

To summarize this part, we underscore that the general isospectral problem for tight-binding hamiltonians of equal sites can be reduced to the isospectral problem of the (locally) periodic chain by means of a family of transformations represented by the matrix $X$. Then by varying $X$ we obtain all the possible orbits of a given matrix $A^{\dagger} A$ using the relation $ X A^{\dagger} A X^{\dagger} = \tilde A^{\dagger} \tilde A \equiv A^{\dagger} A \left[ \alpha \right]$, i.e. a functional of $\alpha$. The final step consists in a reconstruction of $F$ using once more $F_n = |A_n| |A_{n+1}|$, where additional phases are irrelevant due to (\ref{1.9}). The parameter $A_1$ is lost in the expression for $F_n$ (see the power $(-1)^n$ in (\ref{1.18})), and the generation of new couplings $\tilde F_n$ comes exclusively from the variation of all possible $\alpha$'s. We refer once more to figure \ref{fig:1}.

\subsection{Algebraic equations for couplings and energies \label{sec:2.2}}
Here we obtain a set of algebraic equations whose solutions determine the systems that lead to a given (target) spectrum. Our approach in this part is a rather direct one: we shall find equations in the variables $\{ F_n \}_{n=1}^{N+1}$ and parameters $\sigma(H)=\{ E_k \}_{k=1}^{N+1}$.

\begin{figure}[h!]
\begin{center}  \includegraphics[width=7cm, height=7cm]{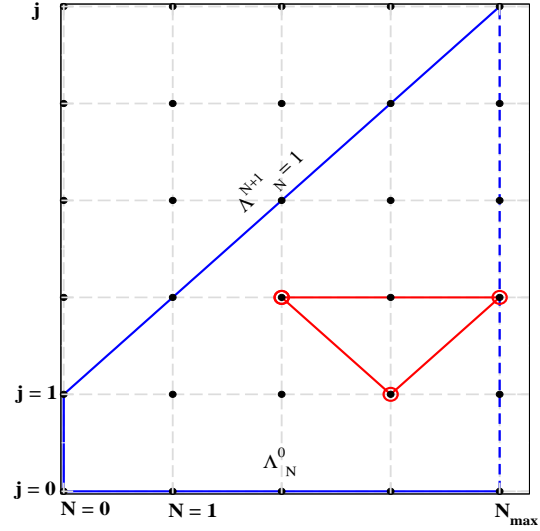} \end{center}
\caption{\label{fig:2} Lattice representation of the recurrence relation (\ref{2.3}) in red and its boundary conditions in blue.}
\end{figure}

We start by defining $\lambda = -E$, $\Phi_N(\lambda) = \mbox{det}\left( H+ \lambda \right)$, leaving us with a secular equation $\Phi_N(\lambda)=0$. An expansion of $\Phi_{N}(\lambda)$ by the method of minors leads to a recurrence relation in the size of the system $N+1$:

\bea
\Phi_N(\lambda) = \lambda \Phi_{N-1}(\lambda) - F_N^2 \Phi_{N-2}(\lambda).
\label{2.1}
\eea
We want to determine the coefficients $\Lambda_{N-1}^{j}$ in the following power expansion

\bea
\Phi_N(\lambda)= \sum_{j=0}^{N+1} \Lambda_N^j \lambda^j.
\label{2.2}
\eea
To this end, we replace (\ref{2.2}) in (\ref{2.1}) and collect terms in $\lambda^j$ to obtain

\bea
\Lambda_{N}^{j} = \Lambda_{N-1}^{j-1} - F_N^2 \Lambda_{N-2}^{j}.
\label{2.3}
\eea
This is a double recurrence in size $N$ and order $j$. In order to solve it, we need to specify boundary conditions on the lattice $(N,j)$ (see figure \ref{fig:2}). The index $j$ is limited by $N+1$, so we have $\Lambda_{N}^{N+1}=1$ (monic polynomial). For $j=0$, we have a lower boundary condition provided by (\ref{2.3}) if we set $\Lambda_{N-1}^{-1}=0$, i.e.

\bea
\Lambda_N^0 = - F_N^2 \Lambda_{N-2}^0
\label{2.4}
\eea
which is solved by 

\bea
\Lambda_{N}^0 = \cases{0&for $N$ even \\ (-1)^{(N-1)/2} \prod_{k=1}^{(N-1)/2} F_{N-2k}^2 & for $N$ odd \\}
\label{2.5}
\eea

It is important to recognize that $N=N_{\mbox{max}}$ in figure \ref{fig:2} is not a boundary condition, but the mere statement that the recurrence shall be solved for a fixed (arbitrary) $N$.

Now we take advantage of the fact that (\ref{2.3}) is of order $1$ in $j$, for the method to solve (\ref{2.3}) consists in iterations of $\Lambda_{N-2}^j$ in terms $\Lambda_{N-3}^j$, $\Lambda_{N-4}^j$ in terms $\Lambda_{N-5}^j$, and so forth. This leads to a relation

\bea
\Lambda_N^j = \sum_{n=0}^{(N-j+1)/2} (-1)^n \Lambda_{N-2n-1}^{j-1} \prod_{l=0}^{n-1} F_{N-2l}^{2}
\label{2.6}
\eea
which puts $\Lambda_{N}^j$ in terms of all the coefficients of a preceding order in $j$, as expected. Now we continue this iterative process by substituting $\Lambda_{N-2n-1}^{j-1}$ in (\ref{2.6}) using (\ref{2.6}) again, but with properly shifted indices. The process ends when we reach $\Lambda_{M}^0$, specified by (\ref{2.5}). The formula thus obtained is

\bea
\Lambda_{N}^{j} = \cases{ 0 & if $N-j$ even \\ \sum_{n_0 = 1, ...., n_{j-1}=1}^{N_{0,j}, ..., N_{j-1,j}} (-1)^{\sum_{i=0}^{j-1} n_i } \prod_{m=0}^j \left\{ \prod_{l_m=0}^{n_m-1} F_{M_m}^2  \right\}   & if $N-j$ odd  \\}
\label{2.7}
\eea
with the shorthands

\bea
N_{k,j} \equiv \frac{N-j+1}{2} - \sum_{i=0}^{k-1} n_i, \qquad 0\leq k \leq j-1,
\label{2.8}
\eea
\bea
M_{m} \equiv N - 2 l_m - 2 \sum_{i=0}^{m-1} n_i - m, \qquad 0\leq m \leq j,
\label{2.8}
\eea
and with $n_m, l_m$ dummy indices. Summations with negative upper limits are simply zero. This general expression for the coefficients of the characteristic polynomial indicates that we are dealing with polynomials of $2N$-th order in the couplings. Also, the coefficients are homogeneous functions of degree $N+1-j$, and setting $\Lambda_{N}^j = \mbox{constant}$ defines a number of $\left[| N/2 |\right]$ surfaces that are shape-invariant under global scaling.

What we have achieved so far is an explicit calculation of the characteristic polynomial for tight-binding hamiltonians of variable couplings. Since we are interested in the solutions of the inverse spectral problem, we proceed now to the determination of $\Phi_N(\lambda)$ in terms of $\{ E_k \}_{k=1}^{N+1}$. We have

\bea
\Phi_N(\lambda) = \prod_{k=1}^{N+1} (\lambda + E_k)
\label{2.9}
\eea
from which the coefficients are obtained by computing the corresponding products: 
\bea
\Lambda_N^j = \sum_{\{ k_i \}} \prod_{r=1}^{N-j} E_{k_r}, \qquad \{ k_i \} \equiv \{ k_i =1, ..., N+1 \forall i; k_i \neq k_j \forall i \neq j\}.
\label{2.10}
\eea
This formula specifies uniquely the characteristic polynomial for a given spectrum; therefore, if we want to find the couplings $F_n$ we must equate (\ref{2.10}) and (\ref{2.7}). The resulting system reads

\bea
\left( \frac{1 - (-)^{N+j}}{2} \right)\sum_{ \{n_k\}}^{\{N_{k,j}\}} (-1)^{\sum_{i=0}^{j-1} n_i } \prod_{m=0}^j \left\{ \prod_{l_m=0}^{n_m-1} F_{M_m}^2  \right\} =  \sum_{\{ k_i \}} \prod_{r=1}^{N-j} E_{k_r}.
\label{2.11}
\eea
This is a set of $\left[| (N+1)/2 |\right]$ equations, giving rise to a number of $N-\left[| (N+1)/2 |\right]=\left[| N/2 |\right]$ free variables and consequently a $\left[| N/2 |\right]$-dimensional surface representing $\scal_H$. We shall see that these relations can be solved uniquely by providing a set of constraints in some of the $F$'s. The numerical method of choice is iterative, but we ellaborate more on this in section \ref{sec:2.5}.

\subsubsection{Minimal conditions for a unique solution} We have seen that specifying the energies leads to a surface $\scal$ of models. It is possible to fix the values of $\left[| N/2 |\right]$ couplings in order to obtain a single point in $\scal$. However, there is another way of defining one and only one model by means of the wavefunction; we show this by considering two cases. If $N$ is even, the size $N+1$ is odd and one of the energies vanishes. Suppose that the corresponding eigenvector has components $\phi_n^{k}$, with $k$ fixed. The relation (\ref{1.7}) implies

\bea
F_n = -\frac{\phi_{n-1}^k}{\phi_{n+1}^k} F_{n-1} =  F_{1} (-1)^{n-1} \prod_{k}  \frac{\phi_{n-1-k}^k}{\phi_{n+1-k}^k} 
\label{2.12}
\eea
fixing all $F$'s up to an overall scale factor $F_1$. If $N$ is odd, we may use our knowledge of any energy $E_k \neq 0$ to write a similar recurrence

\bea
F_n =  \frac{\phi_{n}^k}{\phi_{n+1}^k} E_{k}  -\frac{\phi_{n-1}^k}{\phi_{n+1}^k} F_{n-1} 
\label{2.13}
\eea
which can be solved by iteration. Once the couplings are fixed, the information of other wavefunctions and energies is also fixed.

\subsection{Problems with analytic solutions \label{sec:2.3}}

Quite often one finds that the problem of couplings for a given spectrum is accompanied by additional restrictions. In fact, the latter can make the inverse problem easier to solve, as shown in recent works \cite{sadurni2010-1, sadurni2013, franco2013}.

The aforementioned restrictions may come in the form of Lie algebraic relations satisfied by the hamiltonian or by the operators contained in it. Such relations affect directly the couplings $F_n$ and impose their functional form up to gauge transformations. In table \ref{tab:table1} we give some examples related to Lie algebraic structures imposed as restrictions. The family presented here is related to $\mbox{sl}(2, \mathbb{C})$, which has three generators. This is in fact the most general structure that one may propose, since the hamiltonian $H = FT + (FT)^{\dagger}$ is composed of only two operators $FT, T^{\dagger}F^{\dagger}$ belonging to a Cartan basis. We note however, that the possibilities provided by this well- studied object are sufficiently rich: compact and non-compact subsets of the special linear group give rise to finite or infinite spectra. 
\begin{table}[t]
\caption{\label{tab:table1}%
A class of solvable inverse problems and their algebraic structures
}
\begin{tabular}{lcccc}
\br
\textrm{ -- }  &
\textrm{Dirac oscillator}&
\textrm{Finite oscillator}&
\textrm{Position operator}&
\textrm{Infinite chain}
 \\
\mr
Restrictions & $\begin{array}{c} \{ FT,T^{\dagger}F^{\dagger} \}=H^2, \\  (FT)^2=0  \end{array}$ & $\begin{array}{c} \left[ FT,T^{\dagger}F^{\dagger} \right]= 2 J_z, \\  FT= J_+  \end{array}$ & $\begin{array}{c} \left[ FT,T^{\dagger}F^{\dagger} \right]= \hat N, \\  FT = a  \end{array}$ & $F=1$ \\  &&&&\\
Hamiltonian & $H = \sigma_+ a + \sigma_- a^{\dagger}$ & $H = J_x $ &  $H = x$ & $H = T + T^{\dagger}$ \\
Group & S$(2)$ & SU$(2)$ & Heisenberg & $(\mathbb{Z}, +)$\\
Spectrum & $ \{ \pm \sqrt{n} \}_{n \in \mathbb{N}} $ &  $ \{ -\frac{N}{2}, ..., \frac{N}{2} \}$ & $\Re$ & $\left[ -2,2 \right] \subset \Re$ \\
Coupling $F_n$ & $\sqrt{n}$ or $1$ & $\sqrt{(n-1)(N+1-n)}$ &  $\sqrt{n}$  & $1$ \\
\end{tabular}
\end{table}

\subsection{Exact solutions for short chains \label{sec:2.4}}

 A depiction of the spaces $\scal$ is provided for some examples. The low dimensional cases illustrate in a simple manner the multiplicity of solutions for our inverse problems, which are presented for $N=2,3,4$, as well as the explicit constraints (\ref{2.11}). This renders $\scal$ as a surface or as an intersection of surfaces in the space of couplings. 

\begin{figure}[h!]
\begin{center}  \includegraphics[width=15cm]{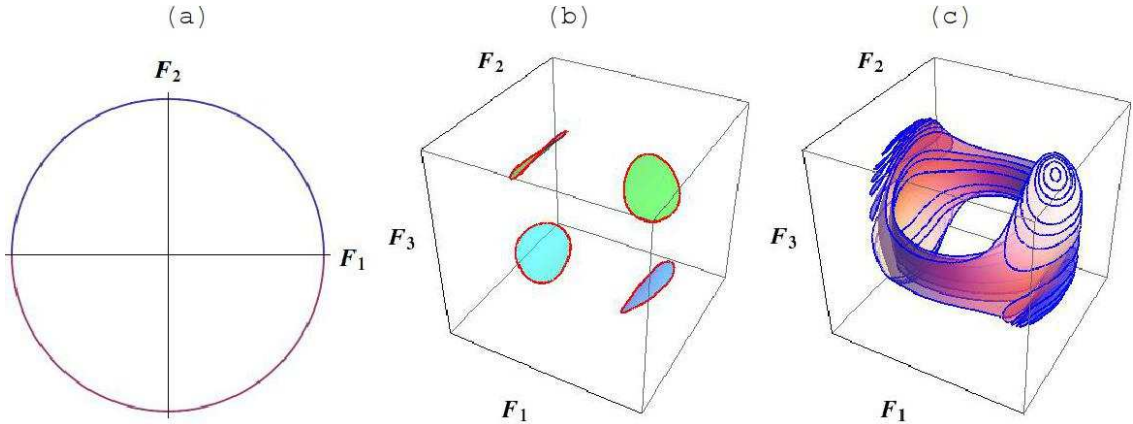} \end{center}
\caption{\label{fig:1.2} Curves and surfaces in the space of couplings $F_1, F_2, ...$, representing isospectral chains. (a) $N=2$, a circle; (b) $N=3$, four loops at the intersections of a sphere with a hyperbolic cylinder; (c) $N=4$, a surface generated by successive intersections of surfaces, each represented by a value of $F_4$.}
\end{figure}
\begin{itemize}

\item[$N=2$:] 

$\sigma(H)= \{ - \sqrt{F_1^2 + F_2^2} , 0 ,  + \sqrt{F_1^2 + F_2^2} \}$ invariant if $F_1^2 + F_2^2=$constant.

\item[$N=3$:]

$\sigma(H)= \{ -E_1 , -E_0 ,+E_0 , +E_1 \}$, with the constraints

\bea
F_1^2 +F_2^2+F_3^2= E_0^2 + E_1^2, \nonumber \\
F_1^2 F_3^2 = E_0^2 E_1^2.
\label{2.14}
\eea
\item[$N=4$:] 

$\sigma(H)= \{ -E_2 , -E_1 ,0, +E_1 , +E_2 \}$, with the constraints
\bea
F_1^2 +F_2^2+F_3^2 + F_4^2= E_1^2 + E_2^2, \nonumber \\
F_1^2 F_3^2 + F_1^2 F_4^2 + F_2^2 F_4^2 = E_1^2 E_2^2.
\label{2.14}
\eea
\end{itemize}

As shown before, in general, adjacent values of $N$ (i.e. with an equal $\left[| (N+1)/2 |\right]$) give rise to sets $\scal$ with the same dimensionality. It is found that $N=1$ yields a circle and $N=2$ yields a loop defined by the intersection of a sphere and a quartic hyperbolic cylinder. The connected components of these surfaces have the same topology. This is displayed in figures \ref{fig:1.2}(a) and (b). When $N=4$, the geometric representation is generated by the successive intersections of two families of surfaces in the space $F_1, F_2, F_3$, giving rise to a familiy of curves parameterized by $F_4$; as a final result a 2d surface in the space $F_1, F_2, F_3$ is obtained. This is shown in figure \ref{fig:1.2}(c).

\begin{figure}[h!]
\begin{center}  \includegraphics[width=12cm]{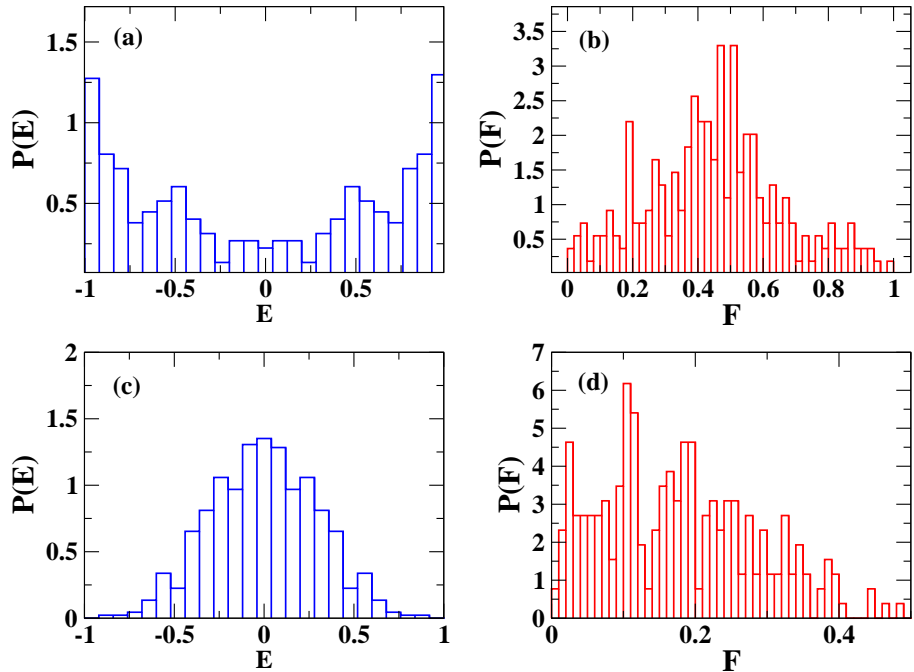} \end{center}
\caption{\label{fig:1.3} Energy and coupling distributions. Upper row: (a) Cosine distribution of energies and (b) peaked distribution of couplings around $F=1/2$ with a nonvanishing width. The peak represents a locally periodic system. Lower row: (c) Gaussian distribution for energies and (d) a distribution of couplings with a smaller variance and non-vanishing average.}
\end{figure}

\subsection{Coupling distributions from random spectra: numerical examples \label{sec:2.5}}

When $N$ is large, the solutions of (\ref{2.11}) can be obtained numerically. As a demonstration, we generate a target spectrum from random distributions and find the family of couplings that produces it. This procedure gives more importance to overall trends rather than specific details. Our goal is to obtain the most probable couplings for two types of spectra: i) a dispersion relation given by a cosine law, emulating a periodic chain and ii) energies of a large system that are located predominantly in a gaussian band with zero average and adjustable width. The exploration of $\scal$ is achieved by choosing random initial conditions in the iterative method (e.g. Newton-Raphson).

 Example i) is relevant to our general theory, given the role of $X$ in (\ref{1.20}). This offers interesting results: If the system is periodic, the spectrum must follow a cosine distribution law, but other possibilities for the couplings may give rise to the same dispersion relation, despite the lack of local periodicity. This is shown in figures \ref{fig:1.3}(a) and (b); in (b) we find indeed a peak at $F=1/2$ as the most problable value (i.e. the periodic case), accompanied by a nonvanishing width that reveals the existence of other systems with the same spectrum (note however that the eigenfunctions do not coincide). The average and deviation are $\< F \> = 0.45$ and $\<\Delta F\>=0.20$ respectively.

The second example gives a distribution of couplings (see histograms in figures \ref{fig:1.3}(c) and (d)) with a decay faster than exponential. This is natural, given the fact that the spectral span is roughly the width $\sigma$ of the gaussian in (c), and the maximum coupling of the system must be constrained also by a gaussian band, i.e. the coupling distribution and the energy distribution have a similar asymptotic behaviour. The average of the coupling distribution is $\< F \>=0.17$ . We also find that the most probable value of the coupling is $F^*=0.11$ and it is considerably greater than zero (here $0 \leq F \leq 0.5$); this is explained by noting that nonzero couplings may give rise to zero or close to zero energy eigenvalues. Thus, if the spectral distribution is dominated by the centre of a band, we obtain matrix elements of $H$ that are predominantly {\it displaced\ }from the centre, as shown in figure \ref{fig:1.3}(d).

\section{Bent waveguide realizations \label{sec:3}}

We test our tight-binding models with a physical system that provides the required flexibility. Adjustable couplings, as well as the existence of evanescent modes draw our attention to systems of waveguides with several bendings. Such realizations are indeed susceptible of manipulations that produce partial isospectrality: we should bear in mind that the Dirichlet boundary value problem posed by bent waveguides leads to nontrivial behavior at high energies or frequencies, connecting the problem with wave-like manifestations of chaos \cite{richter1999}. However, at very low energies we can take advantage of trapped states at corners \cite{bittner2013, exner2015, carini1992, londergan1999, carini1997-1} (in fact, only one bound state at a right bending angle), providing an exponentially evanescent coupling of two corners as a function of the distance between them. The idea then is to accomodate as many levels as necessary below the propagation threshold of the guide; this is done by adjusting the distances between the corners contained in the array. See figures \ref{fig:I.0}, \ref{fig:1.4} and \ref{fig:1.4.1}. In this way we engineer the necessary couplings for a given spectrum and finally tame the beast.

It is important to mention that, from the physical point of view, some realizations put restrictions on the width of waveguides \cite{exner1989-2}. Our realizations are of a different nature, and the existence of bound states in corners is independent of the scale. In fact, they have been confirmed experimentally even with microwaves \cite{bittner2013}. This can be easily explained by noting that the classical theorems found by Exner and Seba rely on the smoothness of the waveguide -- in contrast to our case-- as well as the independent variation of width and curvature. Quoting their result, ``a bound state with energy below the first transversal mode exists for all sufficiently small $d$'', which ensures the existence of a trapped mode for a certain range of the width and ``the proof... establishes existence of the critical width $d_0$...''. In addition, Goldstone and Jaffe (see fig. 1 in \cite{goldstone1992}) proved by a variational method the existence of bound states for all widths in twisted tubes of {\it constant\ } section.
Right-angle bends, on the other hand, have infinite curvature at the corner and do not have constant section. For a single corner, one is convinced that increasing the separation distance between boundaries amounts to a global scaling of the system -- see the explanation below (7) in \cite{sadurni2010-2}. In connection with scaling properties in the limit of vanishing width, some works have shown the persistence of evanescent modes in effectively one-dimensional wires \cite{cacciapuoti2007}, but here we provide an explicit dependence of the evanescence length in terms of the geometry.  

 \subsection{The validity of tight-binding approximations \label{sec:3.1}}

\begin{figure}[h!]
\begin{center}  \includegraphics[width=7cm]{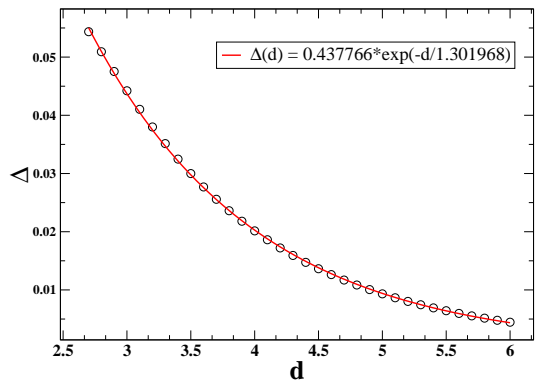} \end{center}
\caption{\label{fig:1.4} Numerical results for the decay of energy splittings as a function of distance $d$ in units of width $L$. The close agreement with the red curve is a clear indicative of exponentially vanishing couplings; the parameters are $\lambda = 1.3 L$, $\Delta(0)=0.43 E_{b}$. This has been computed for U and S configurations shown in figure \ref{fig:1.4.1}.}
\end{figure}

\begin{figure}[h!]
\begin{center} \includegraphics[width=12cm]{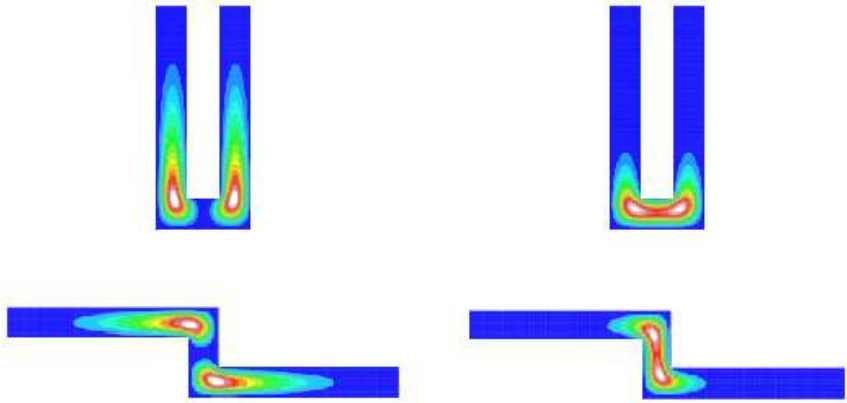} \end{center}
\caption{\label{fig:1.4.1} Symmetric and antisymmetric states for S and U waveguides. The finite separation distance between corners produces a level splitting. }
\end{figure}

The equation to solve here is of the Helmholtz type (a quantum particle trapped in the cavity or TE modes in a good conductor):

\bea
\left[ \nabla^2 + k^2 \right] \phi(x,y) = 0, \quad
\phi(x,y) = 0 \quad \mbox{if} \, (x,y) \in \partial \Omega.
\label{3.1}
\eea
The associated Hamiltonian for this equation is simply $H= -\nabla^2$. For two adjacent corners 1 and 2, we find the coupling $F$ by means of an overlap between trapped modes

\bea
F = \int \int_{\Omega} \phi_1(x,y) H \phi_2(x,y) dxdy,
\label{3.2}
\eea
giving rise to the following block representation of a two-corner system at very low energies:

\bea
H_{\mbox{\scriptsize Low} E} = \left( \begin{array}{cc} E_b & F \\  F  & E_b \end{array} \right).
\label{3.3}
\eea
Here, $k^2 = E_b$ is the energy of a trapped mode in an isolated corner. Numerical calculations show that for an infinite guide $E_b = 0.93 E_t$ \cite{schult1989}, that is, a fraction of the propagation threshold $E_t = \pi^2 / L^2$ if the straight segment has a width $L$. It has been shown theoretically and experimentally \cite{bittner2013} that such trapped modes have exponentially vanishing tails as functions of the distance $d$ from the corner, measured along a straight segment.  We write  $\phi(x,y) \sim \exp \left( - d / 2 \lambda \right)$ with $\lambda \sim 1 / \sqrt{E_b}$. Therefore, the function $F$ can be written to a very good approximation as

\bea
F \equiv \Delta(d) = \Delta(0) \exp \left( -\frac{d}{\lambda} \right).
\label{3.4}
\eea
We have verified this law numerically for a system of two corners in different configurations, the results are shown in figures \ref{fig:1.4} and \ref{fig:1.4.1}. This is also in agreement with the classic results by Carini and co workers \cite{carini1997-1}.

With these considerations we justify the applicability of tight-binding models for the description of spectra and wave functions of more intricate systems of waveguides. We also perform numerical computations of the 2d problem for each realization, in order to give a fair comparison with our predictions.\\

\begin{figure}[t!]
\begin{center}  \includegraphics[width=14cm]{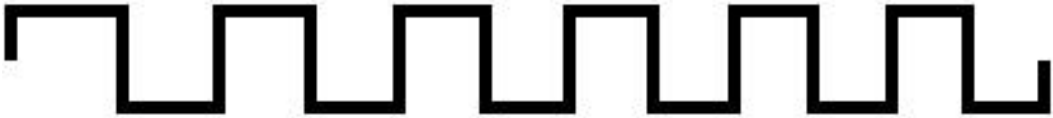} \end{center}
\caption{\label{fig:1.6} A bent waveguide representing a Dirac oscillator. The logarithmically deformed array comprises 11 dimers. Its spectrum corresponds to panel (b) of figure \ref{fig:1.5}.}
\end{figure}

\begin{figure}[h!]
\begin{center}  \includegraphics[width=12cm]{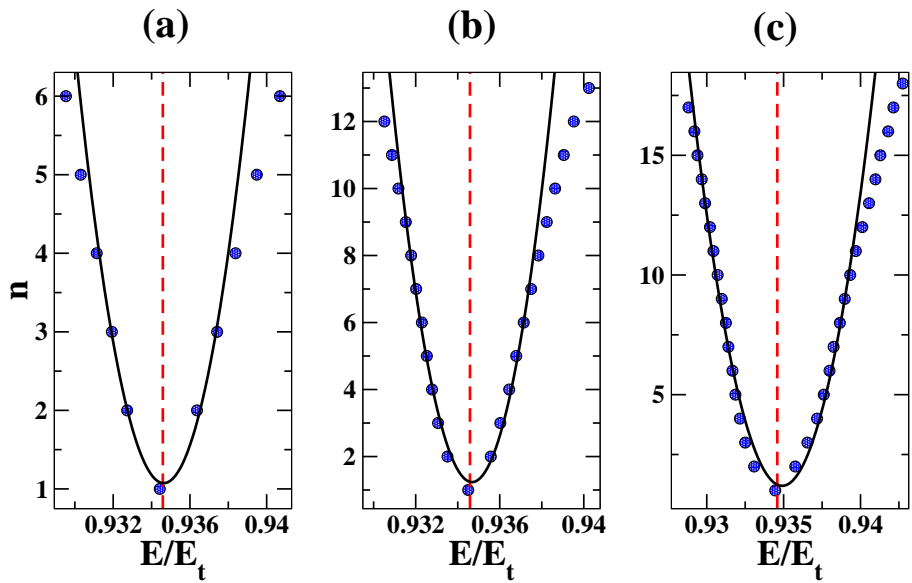} \end{center}
\caption{\label{fig:1.5} Spectra of waveguides representing Dirac oscillators of three different sizes: (a) $N=10$ (b) $N=22$ (c) $N=32$. A better agreement with the theoretical prediction is achieved as the size of the array is increased. More energy levels (blue dots) can be fitted by a square root law (solid black curve) centered at the bound state energy (red dashed line).}
\end{figure}

 \subsection{The realization of exactly solvable configurations \label{sec:3.2}}

Here we provide waveguide designs that produce the models mentioned in table \ref{tab:table1}, as well as the examples discussed in section \ref{sec:2.4}.

    \subsubsection{The Dirac oscillator} This system was recently produced experimentally in \cite{franco2013}, with suggestions provided in \cite{sadurni2010-1}. The incarnation using microwave resonators provided a natural way to adjust couplings by varying distances between dielectric disks. Here we resort to the variation of segment lengths and their orientation to build a sequence of dimers; the intra dimer distance has a constant value in compliance with a constant coupling, while the inter distance of consecutive dimers is varied with a law of the type $\log \sqrt{n}$. For the explicit 1d hamiltonian of this system, see table \ref{tab:table1}. 

The resulting waveguide and the numerically obtained spectrum are shown in figures \ref{fig:1.5}, \ref{fig:1.6}. We observe an excellent agreement with our predictions: a square root law for the spectrum is obtained for half the energies (around the centre of the band). A parabolic curve is fitted to the lowest energies applying the method of least squares to the lower half of the spectrum\footnote{Finite size effects can alter the upper half of the eigenvalues. For this reason we do not include them in the parabolic fit.}.

\begin{figure}[t!]
\begin{center}  \includegraphics[width=13cm]{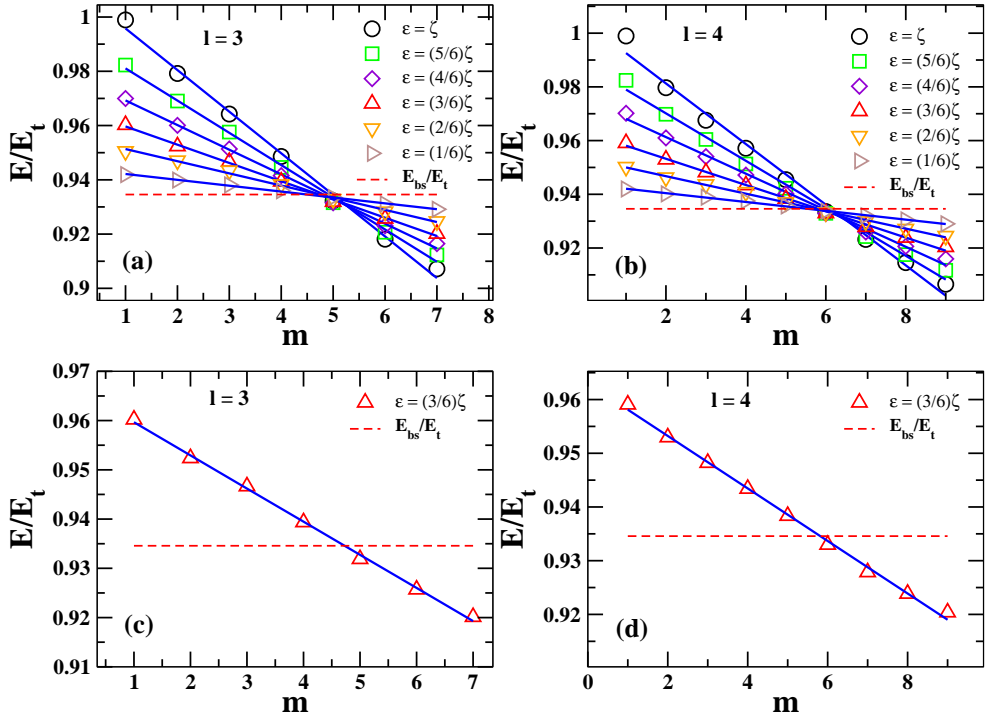} \end{center}
\caption{\label{fig:1.7} Numerical spectra for waveguides obeying the coupling prescriptions of finite oscillators. (a) and (b) show results for various spectral spacings $\varepsilon$ in terms of $\zeta \equiv (E_t-E_b)/ l$ (see inset) and sizes $N=2l=6$ and $8$. (c) and (d) show especial examples where the agreement with equispaced spectra is outstanding.}
\end{figure}

\begin{figure}[h!]
\begin{center}  \includegraphics[width=13cm]{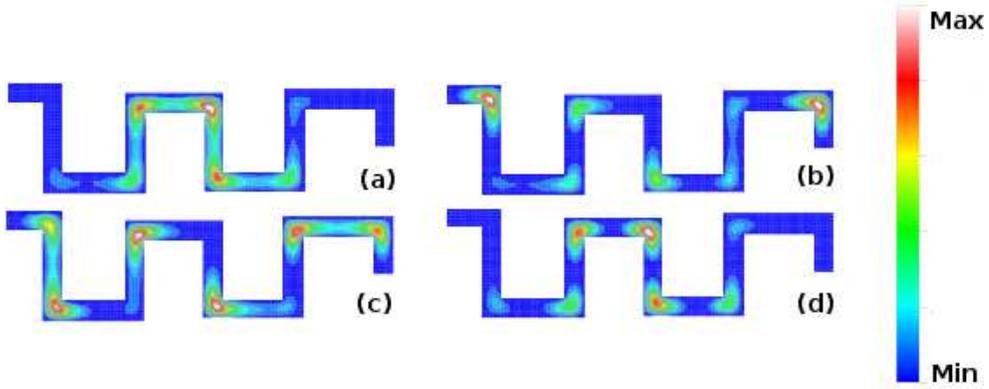} \end{center}
\caption{\label{fig:1.8} Examples of wavefunctions for a finite oscillator with 9 sites, $l=4$. (a) $m=1$, (b) $m=4$, (c) $m=6$, (d) $m=9$. The spectrum is shown in figure \ref{fig:1.7}(d).}
\end{figure}

   \subsubsection{Equispaced spectrum or finite oscillator}

Finite oscillators \cite{atakishiyev1991, atakishiyev2001} and relativistic rotors \cite{sadurni2013} have been proposed in the past, with the aim of producing systems that evolve coherently due to their equispaced frequencies. In one of such applications it was even possible to produce a schematic spectrum for baryons. These constructions do not suffer the effects of finite size, as in the previous case. In its simplest form (see table \ref{tab:table1}), the spectrum corresponds to that of $J_x$, i.e. a projection of angular momentum orthogonal to its quantization axis. Spectra and wavefunctions are presented in figures \ref{fig:1.7} and \ref{fig:1.8}.

\begin{figure}[h!]
\begin{center}  \includegraphics[width=10cm]{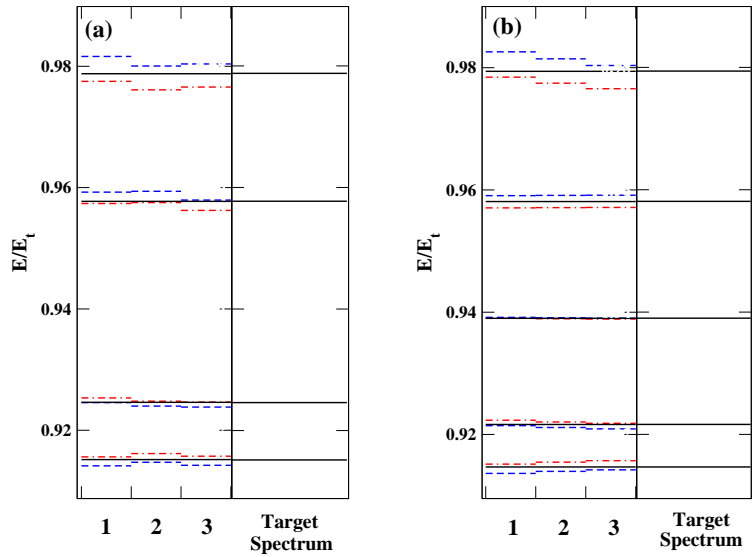} \end{center}
\caption{\label{fig:1.9} The spectra for triads of partially isospectral systems, two sizes: (a) $N=3$ (four sites), (b) $N=4$ (five sites). Abscissa: three systems and a randomly generated target spectrum. Ordinate: energy eigenvalues. The arrays were obtained by finding the couplings and solving for the lengths of the waveguide segments. The boundary value problem was solved numerically and the agreement is quite reasonable. The uncertainties due to discretization were also considered and the corresponding spectrum was calculated when lengths were rounded up (dashed red lines) and rounded down (dashed blue lines). The isospectrality is better achieved numerically in case (c). The arrays are shown in fig. \ref{fig:1.10}}
\end{figure}

\begin{figure}[h!]
\begin{center}  \includegraphics[width=14cm]{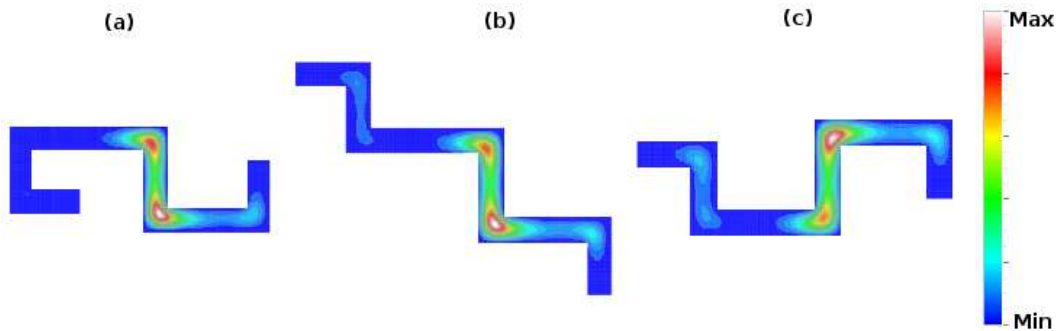} \end{center}
\caption{\label{fig:1.10} Wavefunctions for partially isospectral waveguides. (a), (b) and (c) correspond to systems 1, 2 and 3 in figure \ref{fig:1.9}(b), first excited state.}
\end{figure}

 \subsubsection{Partially isospectral systems for low $N$}

As a final application, we build a family of bent waveguide arrays that possess the same spectrum below propagation threshold. This is most easily done for low values of $N$, already studied in section \ref{sec:2.4}. Three types of waveguides with $N=4, 5$ are designed such that their spectra below threshold coincide. See figures \ref{fig:1.9} and \ref{fig:1.10}. This example is of the utmost importance, since it shows that at least $N+1$ levels of a waveguide (or billiard, if closed at the ends) can be produced by an infinite number of systems. Although the shape of the pipe cannot be determined uniquely using the first $N+1$ levels, we have shown that the set of all pipes can be analytically determined for low $N$.

\section{Conclusion \label{sec:4}}

Inverse lattice design in one dimension shows that arbitrary symmetric spectra can be emulated in various types of tight binding systems. Arbitrary but finite spectra can be also produced by embedding it in half of the energy band obtained from a deformed chain; in this case a bent waveguide operating below threshold was used. The technical goals that we have achieved in this paper are the characterization of the space of all isospectral nearest-neighbour models, the explicit form of the algebraic equations that relate couplings and energies, as well as their analytical (short chains) and numerical (long chains) solutions. Moreover, we designed bent waveguides of variable segments that emulate a number of paradigmatic cases, such as relativistic hamiltonians (a Dirac oscillator) an equispaced spectrum without finite size effects (a finite oscillator) and a few partially isospectral 2d systems (isospectrality only below threshold). In our view, these results are valuable both practically and fundamentally. 
They can be readily used in order to engineer the resonances of many experiments, particularly with microwaves. From the mathematical point of view, we have shown the multivaluedness of solutions and the algorithm to obtain matrix elements from eigenvalues.

\ack

We are grateful to CONACyT for financial support under project CB2012-180585. E.R.-M. also wishes to thank CONACyT for {\it beca-cr\'edito\ } 245104. 

\newpage

\section*{References}


\begin{thebibliography}{99}

\bibitem{band2009}
Band R, Parsanchevski O and Ben-Shach G 2009 {\em J. Phys. A: Math. Theor.\/} {\bf 42} 175202

\bibitem{kac1966}
Kac M 1966 {\em The American Mathematical Monthly\/} {\bf 73} (4) 1--23

\bibitem{berry1989}
Berry 1999 {\em Phys. Scr.\/} {\bf 40} 335

\bibitem{stoeckmann1999}
St\"ockmann H.-J. 1999 {\em Quantum Chaos: An Introduction\/} Cambridge University Press

\bibitem{bloch2005}
Bloch I 2005 {\em Nature Physics \/} {\bf 1} 23

\bibitem{oberthaler1996}
Oberthaler M~K, Abfalterer R, Bernet S, Schmiedmayer J and Zeilinger A 1996
  {\em Phys.\ Rev.\ Lett.\/} {\bf 77} 4980--4983

\bibitem{laurent2007}
Laurent D, Legrand O, Sebbah P, Vanneste C and Mortessagne F 2007
 {\em Phys.\ Rev.\ Lett.\/} {\bf 99} 253902

\bibitem{sadurni2010-1}
Sadurn\'i E, Seligman T H and Mortessagne F 2010
 {\em New J. Phys.\/} {\bf 12} 053014

\bibitem{sadurni2013}
Sadurn\'i E, Franco-Villafa\~ne J A, Kuhl U, Mortessagne F and Seligman T H 2013
 {\em New J. Phys.\/} {\bf 15} 123014

\bibitem{franco2013}
Franco-Villafa\~ne J A, Sadurn\'i E, Barkhofen S, Kuhl U, Mortessagne F and Seligman T H 2013
 {\em Phys. Rev. Lett.\/} {\bf 111} 170405

\bibitem{exner1989}
Exner P, Seba P and Stovicek P 1989
 {\em Czech. J. Phys.\/} {\bf 39} 1181

\bibitem{schult1989}
Shult R L, Ravenhall D G andWyld H W 1998
 {\em Phys. Rev. B\/} {\bf 39} 5476

\bibitem{sadurni2010-2}
Sadurn\'i E and Schleich W P 2010
 {\em AIP Conf. Proc\/} {\bf 1323} 283

\bibitem{bittner2013}
Bittner S, Dietz B, Miski-Oglu M, Richter A, Ripp C, Sadurni E and Schleich W P 2013
 {\em Phys. Rev. E\/} {\bf 87} 042912


\bibitem{kuchment2004}
Kuchment P 2004
 {\em Waves Random Media.\/} {\bf 14} S107--S128

\bibitem{kuchment2005}
Kuchment P 2005
 {\em J. Phys. A: Math. Gen. \/} {\bf 38} 4887--4900

\bibitem{kurasov2005}
Kurasov P and Nowaczyk M 2005
 {\em J. Phys. A: Math. Gen. \/} {\bf 38} 4901--4915

\bibitem{pankrashkin2006}
Pankrashkin K 2006
 {\em Lett. Math. Phys. \/} {\bf 77} 139--154


\bibitem{gesztezy1997}
Gesztezy F and Simon B 1997
{\em J. D'Analyse Math.\/} {\bf 73} 267-297

\bibitem{wannier1937}
Wannier G H 1937
 {\em Phys. Rev.\/} {\bf 52} 191

\bibitem{marzari2012}
Marzari N, Mostofi A A, Yates J R, Souza I and Vanderbilt D 2012
 {\em Rev. Mod. Phys.\/} {\bf 84} 1419

\bibitem{biedenharn1989}
Biedenharn L C 1989
 {\em J. Phys. A: Math. Gen.\/} {\bf 22} L873

\bibitem{macfarlane1989}
Macfarlane A J 1989
 {\em J. Phys. A: Math. Gen.\/} {\bf 22} 4581

\bibitem{manko1993}
Man'ko V I, Marmo G, Solimeno S and Zaccaria F 1993
 {\em Int. J. Mod. Phys. A\/} {\bf 8} 3577--3597


\bibitem{griffiths2000}
Griffiths D J and Steinke C A 2000
 {\em Am. J. Phys.\/} {\bf 69} 137--154

\bibitem{richter1999}
Richter A 1999 {\em Playing Billiards with Microwaves -- Quantum Manifestations of Classical Chaos\/} in Emerging Applications of Number Theory 
(The IMA Volumes in Mathematics and its applications 109) 479--523

\bibitem{exner2015}
Exner P and Kovarik H 2015  {\em Quantum Waveguides\/} (Theoretical and Mathematical Physics) Springer Switzerland


\bibitem{carini1992}
Carini J P, Londergan J T, Mullen K and Murdock D P 1992
 {\em Phys. Rev. B\/} {\bf 46} 15538

\bibitem{londergan1999}
Londergan J T, Carini J P and Murdock D P 1999 {\em Binding and Scattering in Two-Dimensional Systems\/} Springer-Verlag Berlin

\bibitem{carini1997-1}
Carini J P, Londergan J T, Murdock D P, Trinkle D and Yung C S 1997
 {\em Phys. Rev. B\/} {\bf 55} 9842



\bibitem{exner1989-2}
Exner P and Seba P 1989
 {\em J. Math. Phys.\/} {\bf 30} 2574--2580

\bibitem{goldstone1992}
Goldstone J and Jaffe R L 1992
{\em Phys. Rev. B \/} {\bf 45} 100--107

\bibitem{cacciapuoti2007}
Cacciapuoti C and Exner P 2007
{\em J. Phys. A: Math. Theor. \/} {\bf 40} F511-F523

\bibitem{moshinsky1989}
Moshinsky M and Szczepaniak A 1989
 {\em J. Phys. A: Math. Gen.\/} {\bf 22} L817

\bibitem{atakishiyev1991}
Atakishiyev N and Suslov S K 1991
 {\em Theor. Math. Phys.\/} {\bf 85} 442

\bibitem{atakishiyev2001}
Atakishiyev N, Pogosyan G S, Vicent L E and Wolf K B 2001
 {\em J Phys.A: Math. Gen.\/} {\bf 34} 9381




















\end{thebibliography}
\end{document}